\documentstyle[11pt,psfig,epsfig,aaspp4,amsfonts]{article}
\def\degr{\hbox{$^\circ$}}
\def\arcmin{\hbox{$^\prime$}}
\def\arcsec{\hbox{$^{\prime\prime}$}}

\def\fs{\hbox{$.\!\!^{\rm s}$}}

\def\farcm{\hbox{$.\mkern-4mu^\prime$}}
 
\DeclareMathSymbol{\blacksquare}  {\mathord}{AMSa}{"04}
\DeclareMathSymbol{\blacktriangle}      {\mathord}{AMSa}{"4E}
\DeclareMathSymbol{\blacklozenge} {\mathord}{AMSa}{"07}

\begin{document}

\title{Optical follow-up of GRB 970508}
\author{T.J. Galama\altaffilmark{1}, P.J. Groot\altaffilmark{1},
J. van Paradijs\altaffilmark{1,2}, C. Kouveliotou\altaffilmark{3,4},
R.G. Strom\altaffilmark{1,5}, R.A.M.J. Wijers\altaffilmark{6},
N. Tanvir\altaffilmark{6},
J. Bloom\altaffilmark{7}, 
M. Centurion\altaffilmark{8,9},
J. Telting\altaffilmark{9}, R.G.M. Rutten\altaffilmark{9},
P. Smith\altaffilmark{10}, C. Mackey\altaffilmark{10},
S. Smartt\altaffilmark{9}, C. Benn\altaffilmark{9},
J. Heise\altaffilmark{11}, J. in 't Zand\altaffilmark{11}}

\altaffiltext{1}{Astronomical Institute `Anton Pannekoek', University
of Amsterdam,
\& Center for High Energy Astrophysics,
Kruislaan 403, 1098 SJ Amsterdam, The Netherlands}
\altaffiltext{2}{Physics Department, University of Alabama in
Huntsville, Huntsville AL 35899, USA}
\altaffiltext{3}{Universities Space Research Asociation at NASA/MSFC}
\altaffiltext{4}{NASA/MSFC, Code ES-84, Huntsville AL 35812, USA}
\altaffiltext{5}{NFRA, Postbus 2, 7990 AA Dwingeloo, The Netherlands}
\altaffiltext{6}{Institute of Astronomy, Madingley Road, Cambridge, UK}
\altaffiltext{7}{Caltech, Caltech 105-24, Pasadena, CA 91125, USA}
\altaffiltext{8}{Instituto Astrof{\'\i}sica de Canarias, La Laguna,
Tenerife, Spain}
\altaffiltext{9}{ING Telescopes/NFRA, Apartado 321, Sta. Cruz de La
Palma, Tenerife 38780, Spain} 
\altaffiltext{10}{Kitt Peak National Observatory, USA}
\altaffiltext{11}{SRON, Utrecht, The Netherlands}

\begin{abstract}
We report on the results of optical follow-up observations of the counterpart
of GRB 970508, starting 
7 hours after the event. Multi-color U, B, V, R$_{\rm c}$ and 
I$_{\rm c}$ band observations were obtained during the first three
consecutive nights. The counterpart was monitored regularly  in
R$_{\rm c}$ until
$\sim$ 4 months after the burst.  The light curve after
the maximum follows a decline that can be fitted
with a power law with exponent $\alpha$ = --1.141 $\pm$
0.014. Deviations from a smooth power law decay are
moderate (r.m.s. = 0.15 magnitude). We find no
flattening of the light curve at late times.
The optical afterglow 
fluence is a significant fraction, $\sim$ 5\%, of the GRB fluence.
The optical energy distribution 
can be well represented by a power law, the slope of which changed at the 
time of the maximum  (the spectrum became
redder). 
 
\end{abstract}

\keywords{gamma rays: bursts --- gamma rays: individual (GRB
970508)}

\notetoeditor{We wish to include the last figure (PLATE1.ps) as a plate.}

\section{Introduction}
\label{sec:intro}

With the Wide Field Cameras (WFCs; Jager et al. 1995) onboard BeppoSAX (Piro
et al. 1995) it became possible for the first time to rapidly determine 
accurate (3\arcmin\ radius) GRB positions (Costa et al. 1997a,b;
In 't Zand et al. 1997; Heise et al. 1997a); such positions, as it
turned out, are the necessary ingredient for 
succesful counterpart  
searches (see also Takeshima et al. 1997 and Smith et
al. 1997). 
The detection of a fading optical counterpart to GRB 970228 
provided the first arcsecond 
localization of a GRB (Groot 
et al. 1997a;  
Van Paradijs et al. 1997; Galama et al. 1997a,b). Its coincidence 
with a faint extended object (Groot et al.
1997b; Metzger et al. 1997a; Sahu et al. 1997a), which is probably the host 
galaxy, suggests a cosmological distance for the GRB. A second 
optical transient (OT) was found for GRB 970508 (Bond 1997). The measurement  
of a lower bound to the redshift ($z \geq$ 0.835; 
Metzger et al. 1997b) of this optical counterpart
has unambiguously settled the cosmological distance scale to GRB
sources. 

On May 8.904 UT the Gamma-Ray Burst  
Monitor (GRBM) on BeppoSAX recorded a moderately bright GRB
(Costa et al. 1997c), which was also recorded with its Wide 
Field Cameras. The fluences,
recorded with the GRBM (40-700 keV) and WFC (2-26 keV), were 
(1.8 $\pm$ 0.3) $\times 10^{-6}$ erg\,cm$^{-2}$ and (0.7 $\pm$ 0.1)
$\times 10^{-6}$ erg\,cm$^{-2}$, respectively (Piro et al. 1997a).
The  
burst was also recorded (Kouveliotou et al. 1997) with BATSE (Fishman
et al. 1989) with a total 20-1000 keV fluence of
$(3.1 \pm 0.2) \times 10^{-6}$ erg\,cm$^{-2}$, and a peak
flux density (50-300 keV) 
of ($1.66 \pm 0.06) \times 10^{-7}$ erg\,cm$^{-2}$\,s$^{-1}$. 
From optical observations of the WFC error box (Heise et al. 1997b),
made on May 9 and 10, 
Bond (1997) found
a variable object at RA = $06^{\rm h}53^{\rm m}49\fs2$, Dec =
+79\degr16\arcmin19\arcsec (J2000), which showed an increase by $\sim$1
mag in the V band. BeppoSAX Narrow Field
Instrument observations revealed an X-ray transient (Piro et al. 1997b)
whose position is consistent with that of the optical variable. 
Extended emission is not associated with this optical 
counterpart (Fruchter et al. 1997, Sahu et al. 1997b, Pian et al. 1997), and
Natarajan et al. (1997) showed that
either GRB 970508 originated from an 
intrinsically very faint dwarf galaxy or that it occured at a large
distance from a host galaxy ($> 25 h_{70}^{-1}$ kpc). 

We here report on the results of optical photometry of GRB 970508, made
between 0.3 and 110 days after the burst.

\section{Optical Photometry \label{sec:phot}}

We have used the 4.2m William
Herschel Telescope (WHT) at La
Palma to make optical observations in U, B, V, R$_{\rm c}$ and I$_{\rm
c}$ of the WFC error box both with the Prime Focus Camera (PF,
f.o.v. 9\farcm0$\times$9\farcm0), and the Auxiliary Port
Camera (AUX, f.o.v. 0\farcm9 radius). All WHT PF
exposures lasted 600 seconds, except the one made on Aug 26.9 UT which
lasted 1800 seconds. The WHT AUX exposures were 3600 seconds each.
As part of the Kitt Peak
National Observatory Queue observations, observations were also made with 
the 3.5m WIYN telescope, using the WIYN Imager S2KB at the Nasmyth
focus. 
Ten 6\farcm8$\times$6\farcm8 R-band images were
obtained between May 9.15 and 9.23 UT at several 
pointings to cover the whole error box. All exposures lasted 300 seconds.   
A log of 
the observations is shown in Tab. \ref{tab:log}. 

\begin{figure}
\centerline{\epsfig{figure=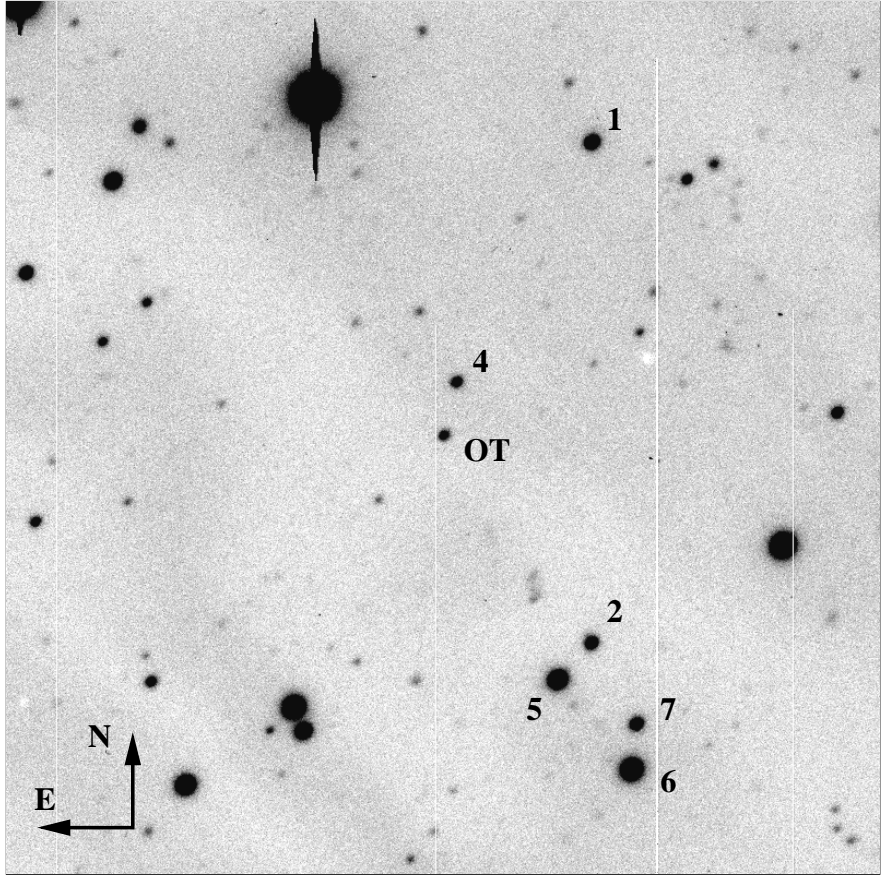,width=12cm}}
\figcaption[]{WHT prime focus 3\farcm5 x 3\farcm5 R$_{\rm c}$ band image of the
field of GRB 970508.  
Indicated are Bond's variable (OT) and the 6 stars used for
calibration. Note that Sokolov's star 3 is not indicated since we did
not use it
as a secondary photometric standard. The U band magnitudes of the four U
band  comparison stars, respectively, stars 2, 5, 6 and 7 are 
19.92$\pm$0.25, 19.26$\pm$0.25, 18.99$\pm$0.25, and 17.61$\pm$0.25.
\label{Rband}}  
\end{figure}

We have corrected the U and B
band images for the non-linearity at low counts of
the LORAL2 CCD Chip\footnote{La Palma webpage http://www.ing.iac.es},
used for  
the May 10, 11, and 12 WHT observations (this effect 
is negligible in the V,  R$_{\rm c}$ 
and I$_{\rm c}$ observations). 
We made a photometric calibration (U, V, R$_{\rm c}$ and I$_{\rm c}$)
using WHT observations on May 9.99 UT  
of the  Landolt (1992) fields PG1047+003 and PG1530+57. 
For V, R$_{\rm c}$ and I$_{\rm c}$ we find good agreement with the results 
of Sokolov et al. (1997). 
To include a B band calibration, and avoid small calibration differences 
with respect to Sokolov et al. (1997)  we used the B, V, R$_{\rm c}$
and I$_{\rm c}$ reference stars (1, 2
and 4 in Fig \ref{Rband}) as given by 
Sokolov et al. (1997) (their star 3 was close to
saturation and therefore not used). 
Stars 2, 5, 6 and 
7 were used as
secondary photometric standards for the U band (see Fig
\ref{Rband}). We have corrected for 
a U band atmospheric extinction of 1.3 times the 
nominal extinction (as Carlsberg Automatic Meridian
Circle 
V band extinction measurements indicate)$^{12}$. 

\section{The R$_{\rm c}$ Band Differential Optical Light curve
\label{lightcurve}} 

\begin{figure}[ht]
\centerline{\psfig{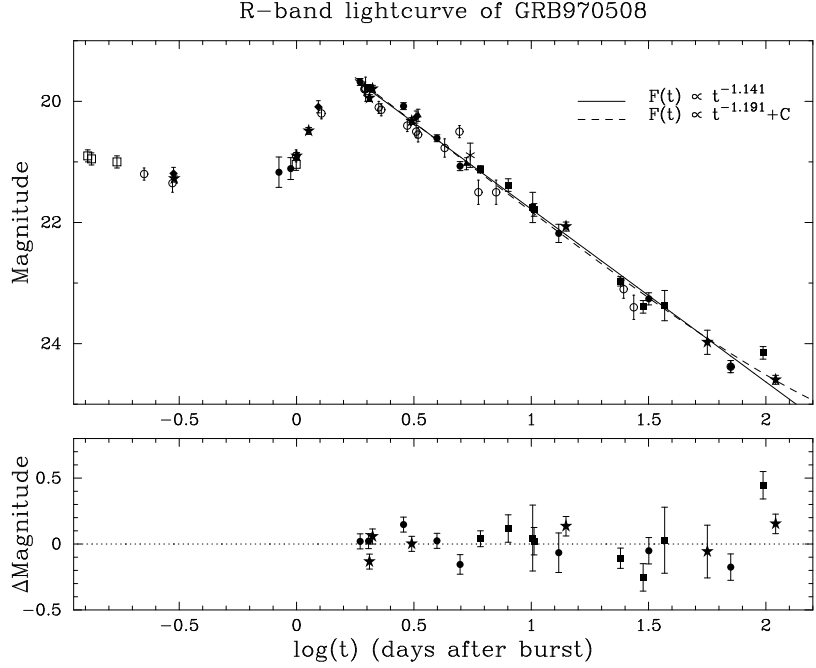}}
\caption[]{Upper panel: the R$_{\rm c}$ band
light curve of 
GRB 970508. For the differential lightcurve (closed symbols) 
data have been used from 
Tab. \ref{tab:log} ($\star$), Sokolov et al. (1997; $\bullet$), 
Pedersen et al. (1997; $\blacksquare$), Garcia et al. (1997;
$\blacktriangle$),
Schaefer et 
al. (1997; $\blacklozenge$) and Chevalier et 
al. (1997; $\times$). The open symbol ($\circ$) represents
non-differential data from Djorgovski et al. (1997), Castro-Tirado et
al. (1998), Mignoli et al. (1997), Kelemen (1997), Fruchter et
al. (1997) and Metzger et al. (1997c). The four open symbols ($\Box$) 
represent wide R filter data by Pedersen et al. (1997).
The power law fit to the second-order corrected data with exponent
$\alpha = -1.141 \pm 0.014$  
is represented by the solid line. Indicated by the dashed line is a
fit to the second-order corrected data of a 
power law plus a constant, $F_{R} = F_{R_{0}} t^{\alpha}+ C$, with exponent 
$\alpha = -1.191 \pm 0.024$. The constant $C$ corresponds to a
magnitude $R_{\rm c} = 26.09 \pm 0.36$ 
Lower panel: the fit to the power law decay subtracted from
the second-order corrected data of Tab. \ref{tab:log} ($\star$),
Sokolov et al. (1997; $\bullet$) and  
Pedersen et al. (1997; $\blacksquare$). 
\label{fig:lightcurve}} 
\end{figure}

In Fig.\ \ref{fig:lightcurve} we present the R$_{\rm c}$ band
optical 
light curve, where we have included results of
differential photometry (relative to local 
reference stars; filled symbols) and calibrated photometry (open symbols).
The differential magnitudes in Tab. \ref{tab:log}. 
were determined with 
respect to stars 1, 2 and 4. The other differential data are 
from Sokolov 
et al. (1997; determined with respect to star 1, 2, 3 and 4), Pedersen et al. 
(1997), Garcia et al. 
(1997), Schaefer et al. (1997) and 
Chevalier et al. (1997). The magnitudes in  
the last four data sets were determined with respect to star 4 only. In the 
absolute calibration of these differential data sets we used the R$_{\rm c}$
magnitudes of these reference stars as 
given by Sokolov et al. (1997). Note that this involves 
a correction to the data of Pedersen et al.
(1997), Schaefer et al. (1997) and Chevalier et
al. (1997) by --0.21 
magnitudes for the difference by that amount in $R_{\rm c}$ 
for reference star 4 used by these authors. 
The difference, in our data, in the R$_{\rm c}$ magnitudes 
of the OT, resulting  
from the use of stars 1, 2 and 4, or only star 4 as a reference star, 
has an r.m.s of 0.027 magnitude. For data sets with a 
sufficiently large 
number of points and covering a time interval of at least 50 days on
the power law decay part of the light curve (i.e., those of Sokolov 
et al. 1997, Pedersen et al. 
1997, and our data) we made a second-order correction by a magnitude
calibration offset, determined from the weighted average offset
with respect to the power law decay part of 
the light curve. The corrections are $0.027 \pm
0.025$, $-0.020 
\pm 0.022$ and $0.082 \pm 0.031$, for, respectively, our data
(Tab. \ref{tab:log}), Sokolov et
al. (1997)  and Pedersen et al. (1997). We have assumed a conservative
minimum error in the differential magnitudes of 0.05 magnitudes.

The R$_{\rm c}$
band light curve
shows a maximum between May 10.2 and 10.8 UT, 1.3--1.9 days after the
$\gamma$-ray burst. We fitted the  
rising and decaying parts of the light curve with power laws, 
$F_{R} \propto t^{\alpha}$. For the fit to the rising part we use both 
the non-differential photometry and the differential data 
(without the second-order correction), assuming a minimum error of 0.10
magnitudes. For the fit to the decaying part we only use the second-order 
corrected data. 
The R$_{\rm c}$ band light curve rises for  
$-0.1 <$ log$(t) < 0.20$ with exponent $\alpha$ = 3.0  
$\pm$ 0.4 ($\chi^{2}_{\rm r}$ = 5.0/6).
After the 
maximum (log$(t) > 0.27$) we find 
$\alpha = -1.141 \pm 0.014$ 
($\chi^{2}_{\rm r}$ = 49/20). 
Including in the fit the
non-differential data and the other differential data, i.e., without the
second-order correction (assuming a minimum error of 0.10
magnitudes) gives $\alpha = -1.132 \pm 0.015$ 
($\chi^{2}_{\rm r}$ = 74.6/38).
Since a hint of flattening of the light curve was reported
(Pedersen et 
al. 1997), we have also fitted a power law plus a constant, $F_{R} =F_{R_{0}}
t^{\alpha} + C$, to the decay part of the second-order corrected 
differential R$_{\rm c}$ band light curve. We find  
$\alpha = -1.191 \pm 0.024$; the constant $C$ corresponds to 
$R_{\rm c} = 26.09 \pm 0.36$ 
($\chi^{2}_{\rm r} = $ 43/19; note that
this corresponds to $R_{\rm c} = 26.30 \pm 0.36$ in the calibration of
Pedersen et al. 1997). 

We also
determined the optical fluence $S$ of the OT (3000 \AA --
10000 \AA) taking for its spectral slope $\beta$ = --1 or --1/3
(see Sect. \ref{sec:spec}). We find $S = (1.20 \pm 0.07) \times
10^{-7}$ erg~cm$^{-2}$ ($\beta=-1$), or $S = (1.90 \pm 0.12) \times
10^{-7}$ erg~cm$^{-2}$ ($\beta=-1/3$) for the fluence collected until
day 110 after the event. Extrapolating the power law decay beyond
day 110 we find a total optical fluence 
$S_{\infty} = (1.36 \pm 0.08) \times 
10^{-7}$ erg~cm$^{-2}$ ($\beta=-1$) and $S_{\infty} =
(2.16 \pm 0.14) \times 
10^{-7}$ erg~cm$^{-2}$ ($\beta=-1/3$).

\section{Broad-band optical flux distribution \label{sec:spec}}

Between May 9.905 UT and May 10.033 UT the R$_{\rm c}$ band flux
increased by 0.137 $\pm$ 0.014 magnitude per hour. In estimating the
spectral energy distribution we corrected
the May 9.9 UT observations for this brightening, i.e., we assumed
that in all passbands the rate of brightening is the same, 
and reduced the magnitudes to a single epoch (May 9.93 UT). 
The May 10.98 UT observations occured at maximum light (see
Fig. \ref{fig:lightcurve}), so we
have not applied any corrections. Corrections to the May 12 data were
calculated using a power law fit to the decay part of the R$_{\rm c}$
band light curve  (see Sect. \ref{lightcurve}). The corrections are 
minor ($< 0.04$ mag) and marginally ($<$ 0.02) affect the spectral slope.  
 
\begin{figure}[ht]
\centerline{\psfig{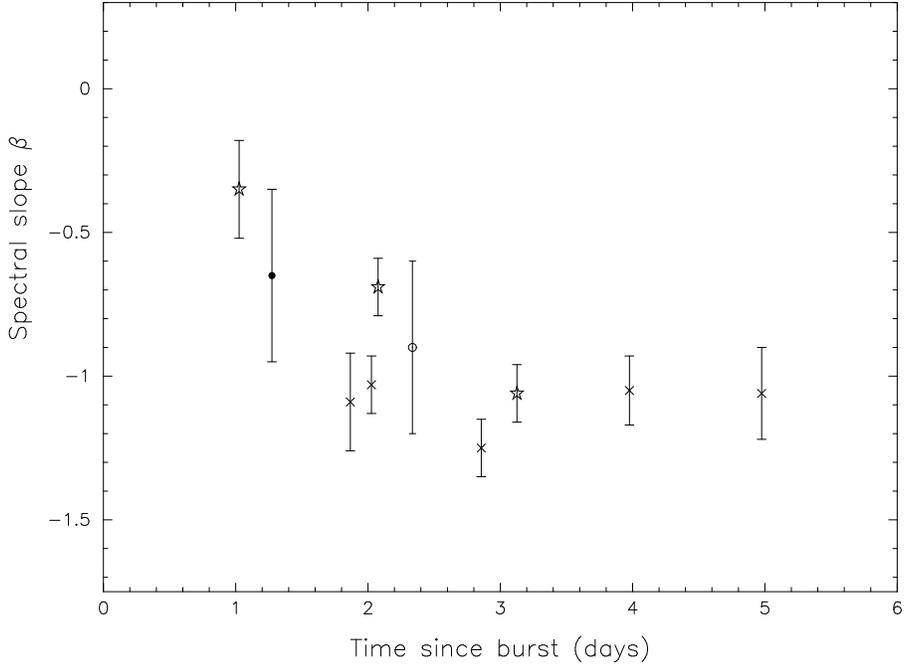}}
\caption[]{The spectral slope $\beta$ as a function of time in
days after the burst (we used $A_V$ = 0.08, see
Sect. \ref{sec:spec}). Data are from: this paper ($\star$), Sokolov et
al. (1997; $\times$), Djorgovski et al. (1997, $\bullet$) and Metzger
et al. (1997b, $\circ$). The maximum of the light curve occured between day
1.3 and 1.9\label{fig:slopes}} 
\end{figure}

We fitted our B, V, R$_{\rm c}$, I$_{\rm c}$ photometry with a
power law, $F_\nu \propto \nu^{\beta}$, using the photometric calibration of
Bessell (1979; table IV) (We chose to exlude the U band data to be able to 
compare our results directly with Sokolov et al., 1997. This exclusion 
does not affect the spectral slope significantly).
Before the fits were made, the data were
corrected for interstellar extinction A$_V$. Using the IRAS
database, we have determined the 100
$\mu$m cirrus flux in the direction of
the OT, $I_{100}$ $<$ 0.08 MJy sr$^{-1}$ 
($2 \sigma)$, corresponding to
$A_V < $ 0.01 (Laureijs et al. 1989).  However,
to compare our results with those of Djorgovski et al. (1997) we also
used their value for the interstellar extinction ($A_V$ = 0.08,
corresponding to their
$A_B$ = 0.11).
The effect of varying the 
interstellar extinction  
on the spectral slope is small; the maximum correction to the exponent $\beta$
(with respect to $A_V$ = 0.08)
is --0.10 (for $A_V$ = 0).  
We have also taken the B, V, R$_{\rm c}$ and I$_{\rm c}$ measurements of
Sokolov et al. (1997) and determined $\beta$,  
correcting for an extinction of $A_V$ = 0.08.
In Figure \ref{fig:slopes} we have plotted  
the spectral 
slopes as a function of time, including the results 
of Djorgovski et al. (1997) and Metzger et al. (1997b).

\section{Discussion \label{sec:dis}}

The differential R$_{\rm c}$ light curve 
shows that deviations from a smooth power law decay are moderate
(r.m.s. = 0.15 mag; lower panel in Fig. \ref{fig:lightcurve}). The
magnitude $R_{\rm c} = 26.09 \pm 
0.36$ for a possible 
underlying galaxy is consistent with the limits derived by Sahu et
al. (1997b) and 
Pian et al. (1997), i.e., $R_{\rm c} > 25.5$. We note, however, that  the
introduction of a constant to the power law fit has not
improved the $\chi^{2}$ substantially; we conclude that the
evidence for a 
flattening of the 
light curve at late times is not yet convincing.

The optical fluence $S_{\infty}$ (3000\AA-10000\AA, $\beta = -1$;
Sect. \ref{sec:spec}), 
is about 5\% of the BATSE 20-1000 keV and 20\% of the WFC 2-26 keV
GRB fluences. A similar comparison with the 
X-ray afterglow (Piro et al. 1997a; $S_{X}$ = 7.3$\times
10^{-7}$ erg\,cm$^{-2}$
2-10 keV, from an extrapolation of 
the X-ray light curve for $t$ $>$ 27 s) shows that the
optical afterglow fluence is $\sim$ 20 \% of the X-ray afterglow fluence.

The optical spectrum of the OT becomes redder before
reaching maximum light (Fig. \ref{fig:slopes}); afterwards the
spectral slope remains constant. We distinguish three phases of the
light curve: phase I, the rising phase ($\alpha =
3.0\pm0.4,\beta=-0.33\pm0.17)$, phase II at maximum light ($\alpha \sim
0,\beta=-0.9\pm0.10)$, and phase III, the decaying phase ($\alpha =
-1.141 \pm 0.014,\beta=-1.11\pm0.06)$. Wijers et al. (1997) give
a relation between the spectral slope $\beta$ and 
the temporal slope $\alpha$ for the simplest fireball remnant model (a
forward blast wave with single Lorentz factor $\Gamma$; their Eq. 2)
and for a `beamed' fireball, i.e. $\Gamma$ and the energy per unit
solid angle E are
functions of 
angle (their 
Eq. 4). For phase III we find very good agreement with $\alpha= -1.141
\pm 0.014$ implying $\beta = -1.094 \pm 0.009$ for the `beamed' case
(marginal agreement with $\beta = -0.761 \pm 0.009$ for the simple case).
Phases I and II are not in agreement with the simple and beamed
fireball models, that predict less steep or even inverted spectral
slopes $\beta$. However, these early phases might be explained by
a continuous distribution of Lorentz factors (Rees \& M\'esz\'aros 1997).
Lower $\Gamma$ (redder) material takes longer to sweep up a
significant amount of 
external matter and hence its emission is delayed with respect to the
higher $\Gamma$ (bluer) material. The lower $\Gamma$ material catches
up gradually with the 
decelerated higher $\Gamma$ material. Hence the rise of the light curve and
evolution from blue to red can be
explained with building up the emission from material with lower
$\Gamma$, while the power law decay (phase III) sets in when 
most of the material has been mixed to a single $\Gamma$.

\section{Acknowledgments}
The WHT is operated on behalf of the English PPARC and Dutch
NFRA at the Spanish observatory Roque de Los Muchachos on La Palma,
Spain. 
The 3.5m WIYN telescope at Kitt Peak National
Observatory, National Optical Astronomy Observatories is
operated by the Association of Universities for Research in Astronomy,
Inc. (AURA) under cooperative agreement with the National Science
Foundation.  The WIYN Observatory is a joint facility of the
University of Wisconsin-Madison, Indiana University, Yale University,
and the National Optical Astronomy Observatories.
T. Galama is supported through a grant by NFRA under contract
781.76.011. C. Kouveliotou acknowledges support from NASA grant NAG
5-2560. 
We thank the BeppoSAX team for the alert.

\newpage

\newpage

\begin{table}
\caption[ ]{Log of observations of GRB 970508 \label{tab:log}}
\begin{tabular}{lllllll}
Date (UT) & Telescope   & U             & B             & V             & R$_{\rm c}$   & I$_{\rm c}$ \\[2mm]
\hline\\[-2mm]
May 9.20  & WIYN        &               &               &               &21.25$\pm$0.05&   \\
May 9.93  & WHT PF      &20.61$\pm$0.25 &               &21.05$\pm$0.05 &20.88$\pm$0.05 &20.49$\pm$0.18 \\
May 10.03 & WHT PF      &               &               &               &20.46$\pm$0.05 &               \\
May 10.98 & WHT PF      &19.75$\pm$0.25 &20.70$\pm$0.08 &20.19$\pm$0.05 &19.92$\pm$0.05 &19.61$\pm$0.30 \\
May 11.01 & WHT PF      &               &               &               &19.77$\pm$0.07 &               \\
May 12.03 & WHT PF      &19.97$\pm$0.25 &21.21$\pm$0.08 &20.71$\pm$0.05 &20.30$\pm$0.07 &19.89$\pm$0.18 \\
May 22.97 & WHT AUX     &               &               &               &22.04$\pm$0.07 &       \\
July 4.19 & WHT AUX     &               &               &               &23.95$\pm$0.20 &       \\
Aug 26.9  & WHT PF      &               &               &               &24.57$\pm$0.07 &       \\
\end{tabular}
\end{table} 

\end{document}